\documentclass{aa}

\usepackage{graphicx}
\usepackage{amsmath,epsfig,rotating,amssymb}

\begin{document}

\def \rot{{\rm {\bf rot} }}
\def \grad{{\rm {\bf grad} }}
\def \div{{\rm div}}
\def \cha{\widehat}
\def \pr{{\it permanent}  regime }


\author{Hennebelle P. \inst{1} and Audit E. \inst{2}}

\institute{ $^1$ Laboratoire de radioastronomie millim{\'e}trique, UMR 8112 du CNRS, 
\newline {\'E}cole normale sup{\'e}rieure et Observatoire de Paris, 24 rue Lhomond,
\newline 75231 Paris cedex 05, France  
\newline $^2$ Service d'Astrophysique, CEA/DSM/DAPNIA/SAp, C. E. Saclay,
\newline F-91191 Gif-sur-Yvette Cedex}

\offprints{E. Audit, P. Hennebelle  \\
{\it e-mail:} edouard.audit@cea.fr, patrick.hennebelle@ens.fr}   

\title{On the structure of the turbulent interstellar atomic hydrogen. I- Physical characteristics}
\subtitle{Influence and nature of turbulence in a thermally bistable flow}

\titlerunning{On the Structure of the atomic hydrogen}

\abstract
{}
{We study in some details the statistical 
properties of the turbulent 2-phase interstellar atomic gas.}
{We present high resolution bidimensional numerical simulations of the 
interstellar atomic hydrogen which  describe it  over 
3 to 4 orders of magnitude in spatial scales.}
{The simulations produce naturally small scale structures having either 
large or small column density. It is tempting to propose that the former 
are connected to the tiny small scale structures observed in the ISM. We compute the mass 
spectrum of CNM structures and find that ${\cal N}(M) dM \propto M ^{-1.7} dM$,
 which is remarkably similar
to the mass spectrum inferred for the CO clumps. 
We propose a theoretical explanation based on a formalism inspired from the 
Press \& Schecter (1974) approach and used the fact that the turbulence within 
WNM is subsonic. This theory predicts ${\cal N}(M) \propto M ^{-5/3}$
in 2D and ${\cal N}(M) \propto M ^{-16/9}$ in 3D. 
We  compute the velocity and the density 
power-spectra and conclude that, although the latter is rather flat, as observed in supersonic 
isothermal simulations, the former follows the Kolmogorov prediction and is dominated 
by its solenoidal component. This is due to the bistable nature of the flow which produces
large density fluctuations even when the rms Mach  number (of WNM) is not large. 
We also find that, whereas the energy at large scales is mainly in the WNM, 
 at smaller scales, it is dominated by  the kinetic energy of the CNM fragments. }
{We find that turbulence in a thermally bistable flow like the atomic interstellar hydrogen,
  is somehow different from turbulence  in a supersonic isothermal gas. 
In a companion paper, we compare the numerical results with atomic hydrogen observations 
and show that the simulations   well reproduce  various observational features. }

\maketitle

\section{Introduction}

The neutral atomic gas (HI) is ubiquitous in the galaxies and is of great importance for 
the star formation process, since molecular clouds form by contraction 
of HI. As such, HI   has been extensively observed over the years 
(e.g.  Kulkarni \& Heiles 1987, Dickey \& Lockmann 1990, 
 Heiles \& Troland 2003, 2005, Miville-Desch\^enes et al. 2003). 
Although HI has also received a lot of attention from the theoretical point of view
(Field 1965, Field et al. 1969, Zel'dovich \& Pikel'ner 1969, Penston \& Brown 1970,
Wolfire et al. 1995, 2003)
and in spite of the  early recognition that HI is a turbulent medium
(e.g. Crovisier 1981, Heiles \& Troland 2005),  it is only recently that the dynamical 
properties of HI have been  investigated.

Two  lines have been simultaneously pursued  by various teams. Attempting 
to understand the properties of HI on large scales, Gazol et al. (2001, 2005),
Dib \& Burkert (2005), de Avillez \& Breitschwerdt (2005), have performed 
2D or 3D numerical simulations of  turbulent multi-phase
 flows subject to various  forcing. 
On smaller scales, various studies attempting to resolve the physical scales 
involved in the problem, have focused on the description of the thermal contraction,
trying to understand  the cooling and the condensation of a piece of WNM into
CNM  accurately.
 Hennebelle \& P\'erault (1999, 2000)  have studied the influence of a converging flow,  
Koyama \& Inutsuka (2000) have investigated the influence of a shock propagating in a 
bistable flow whereas  S\'anchez-Salcedo et al. (2002)  have considered initially 
unstable gas. More recently, Hennebelle \& Passot (2006) have investigated the influence of 
Alfv\'en waves propagating into the medium.
Because of the high  numerical resolution required to treat the problem, 
these works were performed in only one dimension. Further important extensions of these 
works have been performed  in 2 or 3D. 
Koyama \& Inutsuka (2002) have generalized their study in 2D
and Kritsuk \& Norman (2002) have computed the evolution of thermally unstable gas 
using 3D simulations.

In a recent paper (Audit \& Hennebelle 2005 here after paper I), we have investigated the 
dynamical evolution of  colliding  flows of WNM, paying  particular 
attention to the effect of turbulence. The main conclusions of this study 
are as follows: first, we confirm, in a  context where the flow is strongly turbulent, the result
 obtained by Koyama \& Inutsuka (2002)
in their study of shock propagation in WNM.  
The CNM is very fragmented in small clouds which have a dispersion velocity equal
to a fraction of the WNM sound speed. Second, as in Gazol et al. (2001), although at much 
smaller scales, we find 
large fractions of thermally unstable gas and we show that turbulence is able to stabilize
 transiently the thermally unstable gas. Third, we show that in spite of the strong 
turbulence, the CNM fragments are generally not very far from pressure equilibrium with 
the surrounding WNM. Colliding flows of atomic interstellar gas have been further 
investigated by Heitsch et al. (2005, 2006) and by V\'azquez-Semadeni et al. (2006).

One of the crucial issues in modeling atomic hydrogen is the numerical
resolution which is very demanding (see section~\ref{scale}). In this paper, we present 
the result of very high  resolution numerical simulations allowing  description of scales ten 
times smaller than in paper I. We perform a set of simulations with various numerical 
resolutions to study its influence. We also study the influence of thermal conduction by varying 
its value.
The high resolution  allows a  better description of the flow down to 
much smaller scales and offers the possibility to perform various statistical studies
of the flow and CNM structures properties. 

In a companion paper (Hennebelle et al. 2006, paper III), we make preliminary 
comparisons between the  2D simulations presented here and various observational
results. In particular,
the high resolution  permits the comparison with various small scale 
structures that have been observed in HI, like the so-called tiny small atomic structure 
(TSAS) and recent low column density clouds observed by Braun \& Kanekar (2005) and 
Stanimirovi\'c \& Heiles (2005).

In Sect.~2, we describe the numerical setup and method and discuss briefly the 
various scales which have to be adequately treated. In Sect.~3, we present 
one typical timestep of the simulation and discuss qualitatively  various  features of the flow which  
are worth quantifying. Section~4 presents statistical  density and the pressure PDF.  
Section~5 presents  density and velocity power-spectra
as well as the energy spectrum.  
In  Sect.~6, the mass spectrum of the CNM structures in the simulation is computed
and in  Sect.~7, a theoretical explanation is proposed. 
Section~8 concludes the paper.

\section{Initial conditions and method}
\label{condini}

\begin{figure*}
\includegraphics[width=15cm,angle=90]{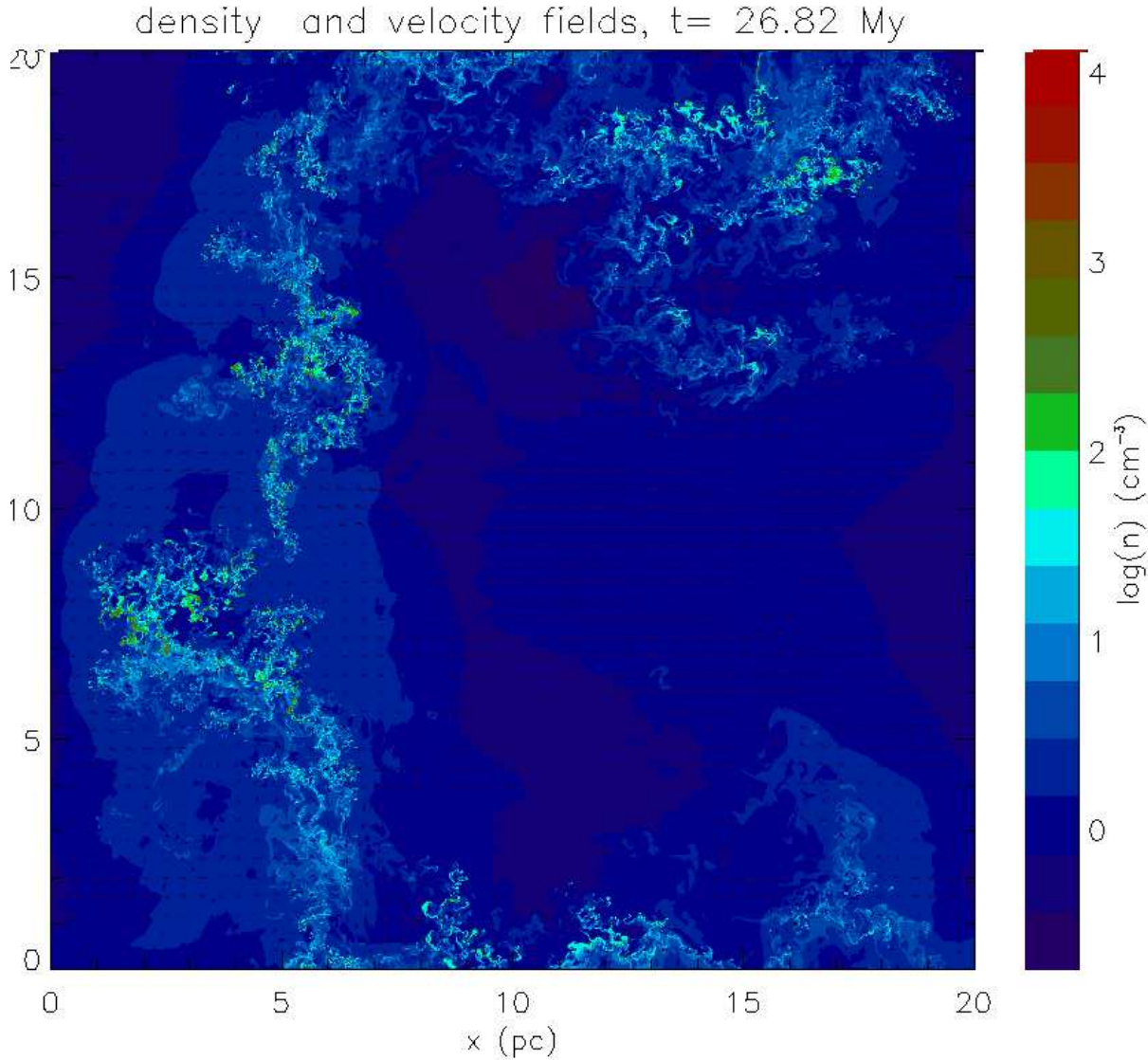}
\caption{Density and velocity fields at one time step of the $10000^2$
cells simulation corresponding to a spatial resolution of $2 \times 10^{-3}$ pc. Note that because of the 
large dynamics, the arrows are not easily seen in the CNM clouds.}
\label{bigchamps}
\end{figure*}

\begin{figure*}
\includegraphics[width=15cm,angle=90]{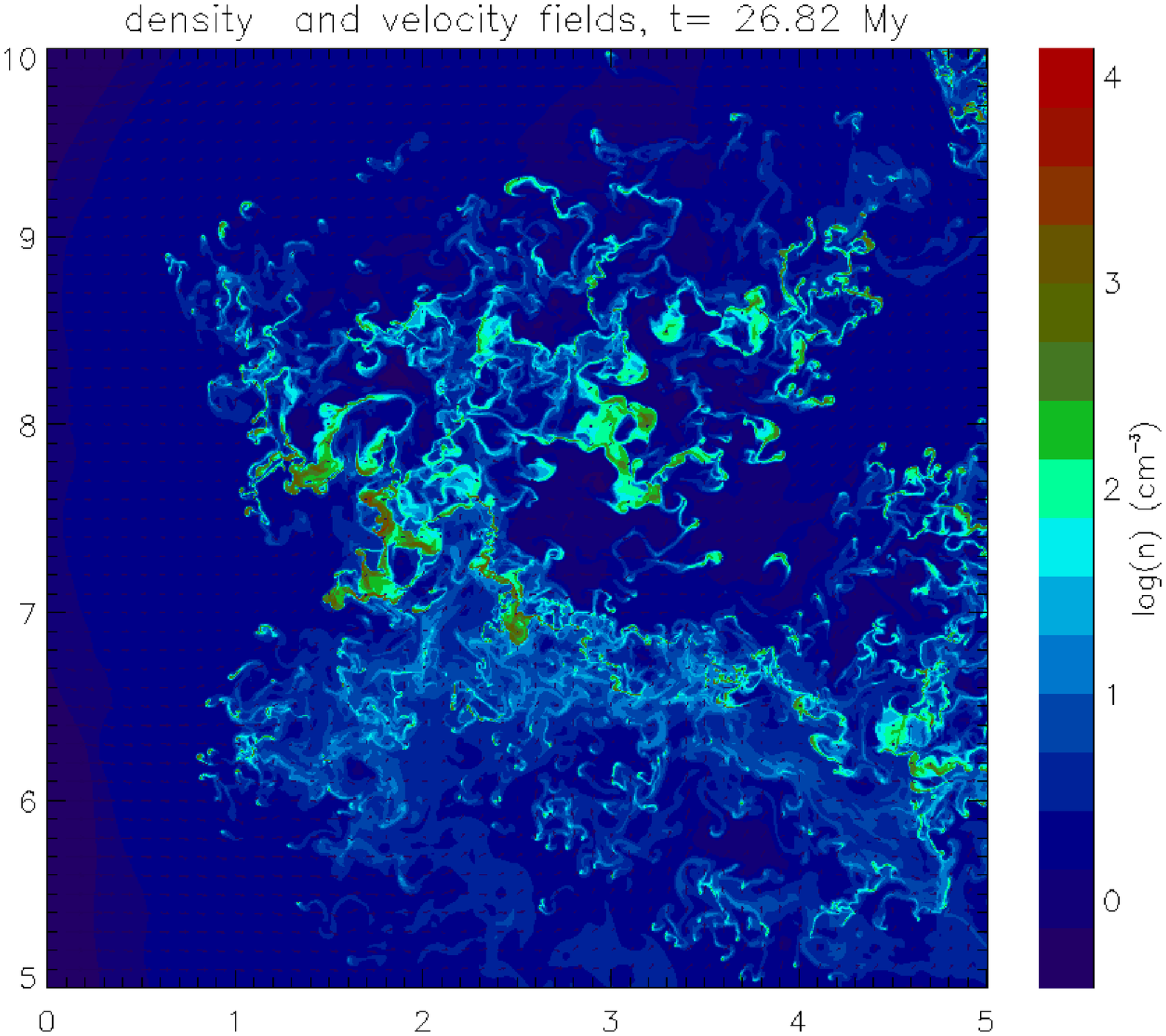}
\caption{Spatial zoom of the field displayed 
in Fig.~\ref{bigchamps}.}
\label{smallchamps}
\end{figure*}

\begin{figure*}
\includegraphics[width=15cm,angle=90]{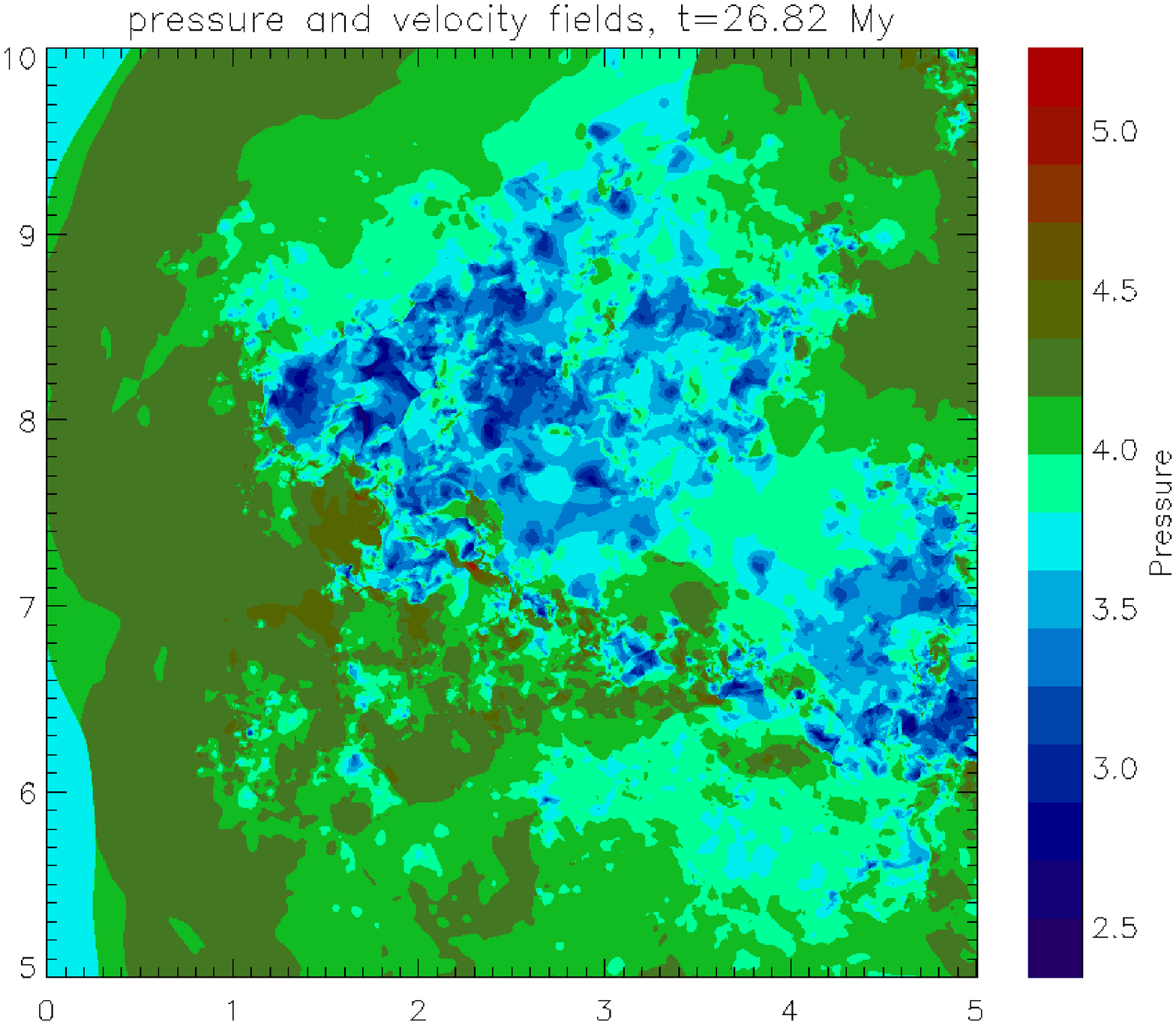}
\caption{Logarithm of the pressure field (in K cm$^{-3}$) corresponding to Fig.~\ref{smallchamps}.}
\label{smallpress}
\end{figure*}

The initial conditions and the methods are very similar to those 
of  paper I. Here, we describe them 
briefly for self-consistency.

\subsection{Equations, notations and numerical method}
We consider the usual fluid equations for a radiatively cooling gas  
including thermal conductivity namely, 
\begin{alignat}{4}
\label{mcons}
\partial_t \rho       & \ +   \nabla . [\rho u]              & =  & 0, \\
\label{momcons}
\partial_t \rho u     & \ +   \nabla . [\rho u\otimes u + P] & =  & 0, \\
\label{econs}
\partial_t  E         & \ +   \nabla . [u(E + P)]            & =  & - {\cal L}(\rho,T) + \nabla ( \kappa(T) \nabla T ).
\end{alignat}
$\rho$ is the mass density, $u$ the  velocity, $P$ the pressure,
$E$ the  total energy and  ${\cal L}$  the  cooling function which 
includes Lyman-$\alpha$,  $C^+$ and $O$ line cooling and grains photoelectric
heating.   
The gas is  assumed to be a perfect  gas with $\gamma = 5/3$
and with a mean molecular weight $\mu = 1.4 m _H$, where $m _H$ is
the mass of the proton.
$\kappa$ is the thermal conductivity and is given by 
$\kappa(T) = \gamma C_v \eta(T)$ where $C_v=k_b / m_H / (\gamma -1)$,
 $\eta=5.7 \times 10 ^ {-5}$ (T/1 K)$^ {1/2}$ g cm$^{-1}$ s$^{-1}$ and $k_b$ is the 
Boltzmann constant.

We use the HERACLES code to perform the simulation. This is a second order
Godunov type hydrodynamical code.  Godunov methods are now widely used and 
tested and are particularly well suited to the treatment of shocks
(e. g. Toro 1997).
The size of the computational domain is $20$ pc and the resolution 
ranges from $600^2$  to $10000^2$ cells leading to a spatial 
resolution ranging from   $3.3 \times 10^{-2}$ pc to $2 \times 10^{-3}$ pc.

\subsection{Initial and boundary conditions}
The boundary conditions consist in an imposed converging flow at the left and
right faces of amplitude $1.5 \times C_{s,wnm}$ on top of which turbulent
fluctuations of amplitude $\epsilon=2$ (see paper I for 
accurate definitions) have been superimposed.
On top and bottom faces, outflow conditions have been setup. This means that 
the flow is free to escape the computational box across these 2 faces.

The initial conditions consist of uniform WNM 
($n \simeq 0.8$ cm$^{-3}$) at thermal  equilibrium. 
It is worth stressing that initially no CNM is present in the computational 
box. The simulations are then  runned until a statistically stationary state 
is reached. Typically, this requires  about 5 to 10 box crossing times.
At this stage, the incoming flow  compensates on average the 
outgoing material.

For the highest resolution runs, this would lead to very long simulations and 
we therefore adopt a slightly different strategy. We start the run with a lower
resolution of $1250^2$ cells and wait until the statistically stationary 
regime is obtained. Then, we increase the resolution of the simulation by a
factor 2 and restart the now 4 times larger simulation. This new simulation
is  runned until the stationary regime is reached. Typically this 
takes about two box crossing time. This procedure is then  repeated until 
the desired spatial resolution is obtained.

\subsection{Scales}
\label{scale}
Contrary to polytropic fluids, a 2-phase medium
has various spatial and temporal characteristic scales which 
have to be adequately described. Here, we briefly recall them.

\subsubsection{A static spatial scale}
The Field length (Field 1965) is the length at which thermal 
diffusivity becomes comparable to the heating and cooling 
term. Its typical value is about 0.1 pc in the WNM and 
about $10 ^{-3}$ pc in the CNM. The Field length is also 
the typical scale of the thermal fronts that connects the cold and warm phases.
As demonstrated by Field (1965), it determines the smallest wavelength at 
which thermal instability can grow and, therefore, 
 the size of the smallest CNM structures. 
The effect of varying the thermal diffusivity is discussed in the following.  

\subsubsection{Three dynamical scales}
We distinguish three scales whose origin is due to dynamical 
processes. The first scale is the cooling length of the WNM, i.e. the product 
of the cooling time and the sound speed within WNM, 
$\lambda_{\rm cool}= \tau_{\rm cool} \times C_{\rm s,WNM}$. This scale is about 10 pc
and corresponds to the typical length at which WNM is non linearly unstable
and can be dynamically triggered by a compression into the unstable regime.

Since the ratio between the CNM and WNM  density is about hundred, the
size of a CNM  structure formed through the  contraction of a piece of
WNM of  size  $\lambda_{\rm  cool}$,  will be  typically, hundred  time
smaller (assuming a monodimensional  compression). Therefore, the size
of the CNM structures is  about the cooling  length of WNM divided  by
hundred, leading to about $\lambda_{\rm cool}/100 \simeq $0.1 pc
\footnote{In fact, as we will see in the following, the size of the CNM structures follows a 
density distribution. The numbers given here are simply indicative}.

Finally, as first pointed out by Koyama \& Inutsuka (2002), the fragments
of CNM have a velocity dispersion equal to  a 
fraction of the sound speed of the medium in which they are embedded, i.e. 
WNM. Since the contrast between the sound speed within the two phases is about 
10, it means that CNM structures undergo collisions at Mach number, 
${\cal M} \simeq$  10. 
If, for simplicity, we assume isothermality and apply  
Rankine-Hugoniot relations, we obtain that the size of the shocked 
CNM structures is given by the size of the CNM structures divided by 
${\cal M}^2$ which is about $10^{-3}$ pc, whereas the density peak is given by 
$100 \times {\cal M}^2 \simeq 10^4$ cm$^{-3}$.
Note that since the effective polytropic index of the CNM,
 $\gamma _ {eff}$, is closer to 0.7 than to 1, these numbers are underestimated by a 
factor of about 7. 

The consequences of not solving properly these scales have been previously 
discussed qualitatively in paper I.  In the following,  quantitative estimates are given.

\subsection{Comparison with other simulations}
As explained in the previous section, it is actually not possible to 
treat simultaneously all the scales involved in the physics of the 
atomic gas. This implies  compromises and induces various choices
of initial, boundary and forcing conditions. In order to make clear 
the domain of validity of various related works, we now discuss the numerical 
and physical setups which have been chosen. It is worth stressing that 
the different choices are complementary and equally interesting since 
they  allow one to focus on different aspects of the same physical problem.

Koyama \& Inutsuka (2002) have investigated the propagation of a shock 
into WNM. They have an extremely high spatial resolution ($7 \times 10^{-4}$ pc) and a relatively 
modest box size of about 0.25 pc (in the transverse direction). With these choices, 
the CNM structures are very well resolved but the statistics are low and the structures
are small. In a recent paper (Koyama \& Inutsuka 2006), they consider a larger box and 
try to understand the properties
of the turbulence that develops without any external forcing.

Heitsch et al. (2005, 2006) and V\'azquez-Semadeni et al. (2006) have focused 
on the formation of a CNM turbulent layer. They consider a simulation box of 
about $\simeq$10 pc and have a resolution of about $ 10^{-2}$ pc. Such a setup 
which is very similar to the one used in paper I, allows to 
treat reasonably well the large scales, but does not provide a  description 
of the CNM structures as accurate as the CNM structures obtained in 
Koyama \& Inutsuka (2002). Also in these work, the authors focus on the 
transient regime,  rather than on a time-independent situation. 

Gazol et al. (2001, 2005) and Piontek \& Ostriker (2004)
 have focused on larger simulation boxes (100 pc-1 kpc)
and coarser spatial resolution (0.5-5 pc). The same kind of scales has been investigated 
by De avillez \& Breitschwerdt (2005) using an AMR scheme. 
This leads to a much better description
of the larger scales and a more realistic self-consistent forcing 
(e.g. due to supernovae or the development of the magneto-rotational instability)
that the choice makes in this work, but 
only marginal description of CNM structures can be achieved. 

\section{Results}
In this section, we present results for the $10000^2$ cells simulation
which corresponds to  the highest numerical resolution runs.
 We discuss qualitatively the main properties 
of the 2-phase turbulent flow and introduce the most important aspects
which are studied quantitatively in the following and in paper III.

\subsection{Main features}

\begin{figure*}
\includegraphics[width=15cm,angle=90]{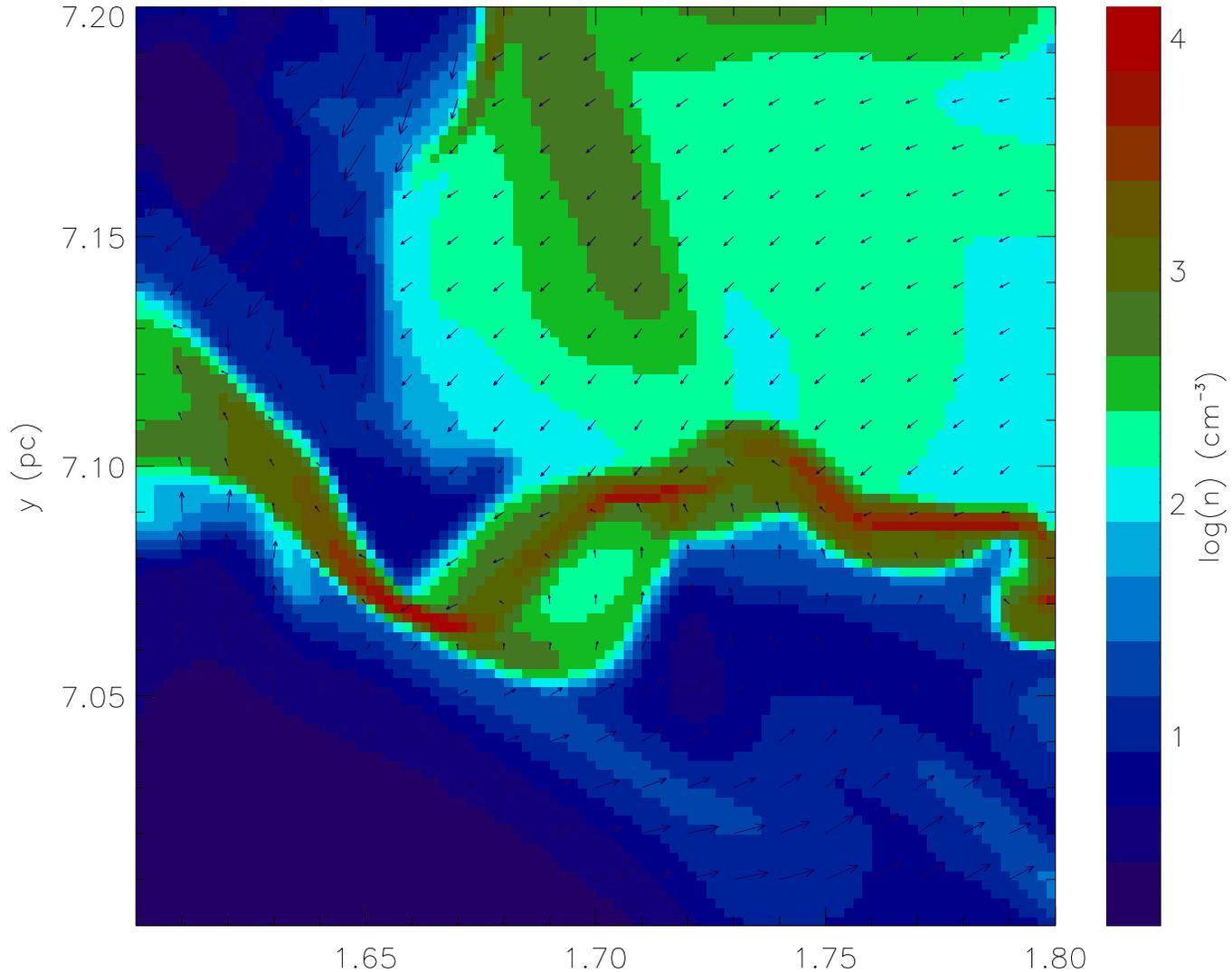}
\caption{Spatial zoom of the fields displayed 
in Fig.~\ref{bigchamps}.}
\label{smallchamps2}
\end{figure*}

Figure~\ref{bigchamps} displays the density and velocity fields of the 
simulation at one representative timestep (i.e. well after statistically stationary state
has been reached)  whereas Fig.~\ref{smallchamps}
 shows a  spatial zoom of 5 pc.

With our choice of  boundary conditions which entails a  converging flow,
  the formation of a layer-like structure is triggered in the middle of the 
computational  box  ($x \simeq 5$ pc, $y \simeq 2-18$ pc).  

The fragmentation of such a layer due to the 
development of various instabilities, like the Vichniac instability
(Vichniac 1994) 
or the Kelvin-Helmholtz instability has been studied by Walder \& Folini (1996, 1998)
and  Heitsch et al.  (2005, 2006) for a radiatively cooling gas 
and by Folini \& Walder (2006) for strongly supersonic  isothermal gas.  
Here, since the converging flow imposed at the boundary presents 
strong velocity fluctuations, the  turbulence and the fragmentation of the 
layer are not easily attributable to a single instability and at least, partly 
due to the driven turbulence. Indeed, in paper I, we vary the amplitude of 
the shear velocity fluctuations which are superimposed to the converging flow. 
As expected, we find that the stronger these fluctuations, the stronger the 
turbulence in the computational domain.
Near the boundaries of the computational box, the 
structure of the flow is somehow different because the converging flow
 is concentrated in the middle of the computational box faces, 
whereas the ram pressure at the edge of the box faces almost vanishes.

 The overall structure of the flow appears to be
highly complex. The cold phase is  very fragmented and the two phases are highly
intervowen. 
There are large regions of space where there is only WNM and 
regions where most of the mass is  obviously in the CNM. However, even 
in these denser regions, WNM is still present   and intrusive.
Only few large CNM structures have formed, most of them appear to be 
very small. This is more clearly seen in Fig.~\ref{smallchamps}
which also reveals that the structures are often connected to each 
other by filaments of lower density and that the density of the CNM varies 
over 2 orders of magnitudes. 

\subsection{Shocked CNM: TSAS ?}
While most  of the CNM structures have a  density of about $100$ cm$^{-3}$ 
(e.g. structure located at $(x,y) = (3,8)$ pc), 
shocked regions present densities up to $\simeq 10^4$ cm$^{-3}$ (e.g. structure located 
at $(x,y)=(1.66,7.07)$ pc).
This is more clearly seen in Fig.~\ref{smallchamps2} which shows a spatial zoom 
of Fig.~\ref{smallchamps} and in Fig.~\ref{coupe} where a cut along the y-axis is 
displayed. These two figures also reveal that the density peak corresponds to a peak 
of pressure reaching a value of about $10^5$ K cm$^{-3}$ and that this strong 
fluctuation has been induced by a converging flow of velocity $\pm 5$ km/s. 
Altogether, this is perfectly consistent with the orders of magnitude given in
paper I and recalled in Sect.~\ref{scale}
 to characterize the  shocked CNM structures.
The density ($\simeq 10^4$ cm$^{-3}$) and the size ($\simeq 400$ AU) appear to be close
to what is inferred for the TSAS structures observed in the CNM (e.g. Heiles 1997), 
although the more extreme events present densities up to 10 times this value 
and scale roughly 10 times smaller. 
It is therefore tempting to propose that TSAS are produced by supersonic collisions 
between CNM fragments. It should be noted that the density peak is not fully resolved 
because of insufficient numerical resolution and therefore it is expected that this 
event should have indeed a larger peak density and a smaller size if the
 numerical resolution was higher. 
It is also important to emphasize that the cooling function used
in this simulation is probably not accurate for such high densities.

\subsection{Small column density clouds}
In a similar way,  the properties
of the smallest  structures that form in the simulations appear to be reminiscent 
of the tiny structures which have been recently observed by 
Braun \& Kanekar (2005) and Stanimirovi\'c \& Heiles (2005). 
  Many of them  can be seen in Fig.~\ref{smallchamps} (e.g. for $y \le 6$ pc)
and a cut through two small CNM structures is displayed in  Fig.~\ref{coupe} 
($y \simeq 6.68$ and 6.83 pc). The density of these two structures is about 
30 and 100 cm$^{-3}$ whereas their size is about 10$^{-2}$ pc.

Further examples are shown in paper III in which synthetic HI spectra
are also discussed.

\subsection{Pressure distribution}
Another important aspect  is revealed by the pressure field 
displayed in Fig.~\ref{smallpress}. The pressure field appears  well organized 
 with large scale systematic gradients. At large scales, the pressure and the 
density fields appear to be broadly anti-correlated. Indeed, the central cloud
(located roughly between $x=1$ and $x=4$ and $y=6$ and $y=9$) is surrounded by  
high pressure WNM, whereas its internal pressure is a few times lower. This is 
a consequence of the radiative cooling. When the WNM fluid particles enter into the 
shocked layer, their internal pressure goes up like in an adiabatic shock, then 
they cool and finally condense out. Indeed, the cooling length of 
shocked (Mach $1.5-2$) WNM is few parsec which is the length of the 
high pressure WNM layer seen on the left part of the box. Interestingly, the 
high density shocked CNM is located at the interface between the high pressure
WNM and the low pressure cloud. Note that since  the WNM pressure ($P \simeq 
10^4$ K cm$^{-3}$) is lower by 
at least one order of magnitude than  the large pressure fluctuation ($P \simeq 
10^5$ K cm$^{-3}$), it cannot be the triggering agent. 
However, this high pressure WNM is
 in part responsible of these large pressure fluctuations since it actively induces  the formation of 
relatively high velocity CNM structures, which  then undergo supersonic collisions.

\begin{figure}
\includegraphics[width=9cm,angle=0]{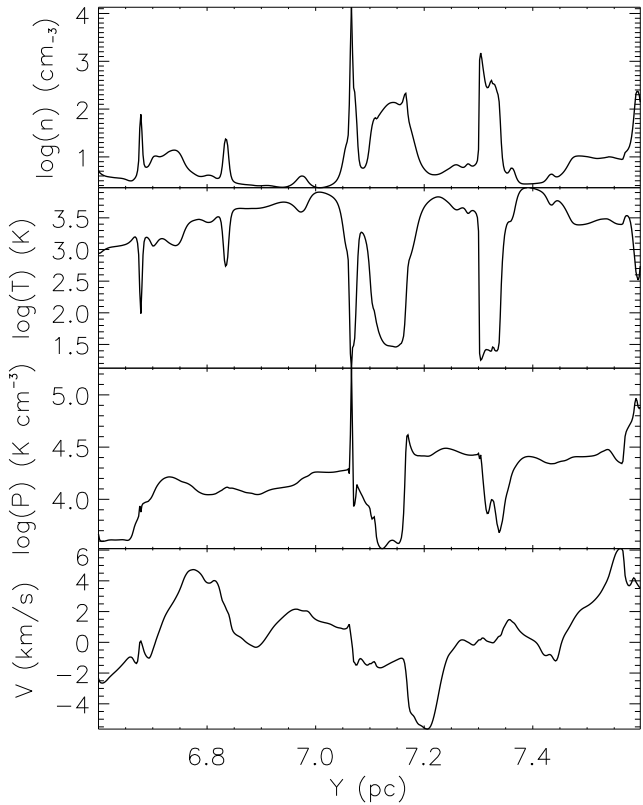}
\caption{Density, pressure and velocity profiles at $x=1.65$ pc for y ranging from 
6.6 to 7.6 pc. Note that this corresponds to the line of sight which intercepts
the cell having the highest density at this timestep.}
\label{coupe}
\end{figure}

At small scales, the situation is much different. The pressure  is poorly correlated 
with the density field displayed in Fig.~\ref{smallchamps}. 
Most of the CNM structures do not have high thermal 
pressure  (see  Fig.~\ref{coupe})
with the notable exception of the densest regions which correspond to  
shocked CNM. Indeed, some of the CNM structures have a thermal pressure lower
than the surrounding WNM as for example, the 2 structures located at $y=7.15$ and 7.3 pc
in Fig.~\ref{coupe}. This low pressure indicates that the structures have just formed,  so that 
there was not enough time for the initially low  thermal pressure induced by the radiative cooling
to readjust to the surrounding higher pressure.

Finally, the density field indicates   
 that the CNM structures are well identified and 
connected to the surrounding WNM by very stiff fronts. 
Moreover, as in the static 2-phase medium (Field et al. 1969, 
Wolfire et al. 1995), most of the CNM structures are not far from quasi-pressure 
equilibrium with the surrounding medium (in the sense that pressure fluctuations are small 
with respect to the density contrast between the two phases).
This is likely a consequence of the large density contrast between the phases. 
First, with a  density contrast of about hundred, the sonic waves propagating 
in the WNM as well as the eddies get mostly reflected when 
they reach a WNM/CNM interface,  probably making  the energy injection inside 
the structures relatively inefficient. 
Second, since the scale of the CNM 
structures is small compared to the scale of the WNM flow and since 
most of the turbulent energy is on the largest scales, the WNM flow
is relatively uniform at the scale of the CNM structures and 
mostly advects them uniformly.

To summarize, it seems that  turbulent 2-phase flows cannot 
be accurately described as a polytropic turbulent flow, neither as 
a static 2-phase medium,  but, instead, they present a kind of duality, 
some aspects being more reminiscent of turbulent flows and other
of static 2-phase media with fluctuations which  can be 
up to hundred times the mean CNM density or pressure.
In the following, we  therefore quantify the mass distribution of CNM structures, as 
well as the density and pressure distribution of the flow.
We  also study the statistics of the flow by computing  various power-spectra,
 in order  to characterize its  scale dependence.
 In paper III, we quantify further the numerical results by computing
line of sight  statistics and various properties  of the CNM structures.

\section{Density and pressure probability distribution function}

\begin{figure}
\includegraphics[width=9cm,angle=0]{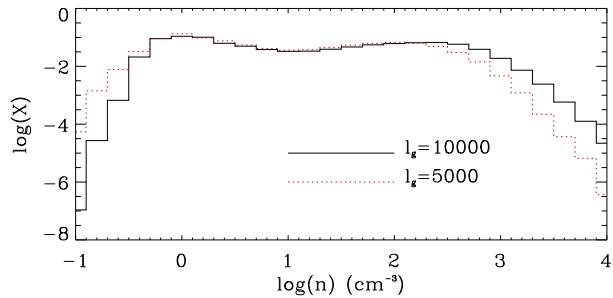}
\caption{Probability distribution function of the logarithm of the density field.}
\label{pdf_dens}
\end{figure}

\begin{figure}
\includegraphics[width=9cm,angle=0]{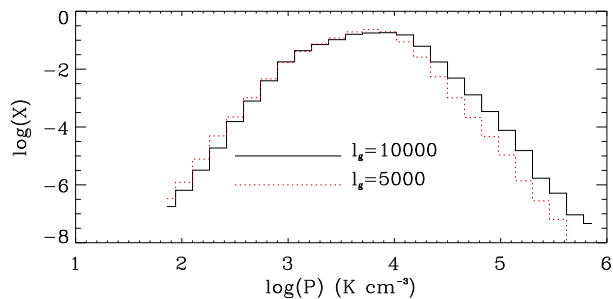}
\caption{Probability distribution function of the logarithm of the pressure field.}
\label{pdf_press}
\end{figure}

Figures~\ref{pdf_dens} and~\ref{pdf_press}  show 
the probability distribution function of the logarithm of 
the density and pressure respectively. 
Full black lines correspond to the runs having $l_g^2=10000^2$ cells,
 whereas the dotted red lines are for the  runs with $l_g^2=5000^2$ cells.
The density pdf presents two peaks which correspond 
to the WNM and CNM densities. The pdf values at densities between 
5 and 50 indicates that there is a significant fraction of thermally unstable
 gas, as first stated by Gazol et al. (2001) and confirmed in paper I.  

In principle, the density pdf allows  estimation of the fraction of the dense gas that
has been  compressed by the ram pressure. However it is seen that the results for the 
high density part do clearly depend on the numerical resolution, meaning that 
higher resolution runs will lead to a higher fraction of high density gas. It should 
also be clear that since the fraction of high density gas depends on the cooling 
processes, it will certainly be modified by a more accurate treatment of the 
microphysics. Having these limitations in mind, we can indicatively say that the 
fraction of gas denser than $\simeq 10^3$ cm$^{-3}$ is about 1-3 $\%$ for this 
simulation. It is important to keep in mind that, 
as emphasized in paper I and in Gazol et al. (2005), this  obviously depends on the 
driving of the turbulence.

The pressure pdf indicates that the average pressure is about $6000$ K cm$^{-3}$,
which is the pressure of the WNM injected at the boundary. The large pressures
 are induced by the ram pressure of the incoming flow. The pressure distribution 
is reminiscent of pdf found by various authors in different contexts (Passot \& V\'azquez-Semadeni
1998, Scalo et al. 1998). Comparison with the results presented by Gazol et al. (2005) is more 
straightforward since they consider a thermally bistable flow as well. As in their case,  
we find that a near-powerlaw tail develops at high and low pressure 
(in their case they report that this depends on the driving length). Because of the higher 
numerical resolution, we reach higher values of pressure. Our pressure pdf appears 
to be somehow similar to the results presented in their figure 11 corresponding to a
large scale driving. However, the significant difference between the 2 numerical 
experiments precludes detailed comparisons.

\section{Velocity and density power-spectra and energy  spectrum}

\begin{figure}
\includegraphics[width=9cm,angle=0]{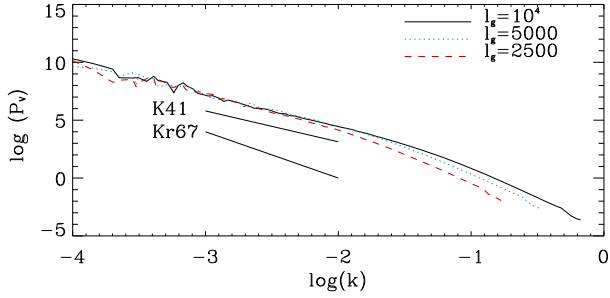}
\caption{Velocity power-spectrum for  numerical resolutions equal to 
$dx=$ $8 \times 10^{-3}$, $4 \times 10^{-3}$ and $2 \times 10^{-3}$ pc.}
\label{velo_ps}
\end{figure}

\begin{figure}
\includegraphics[width=9cm,angle=0]{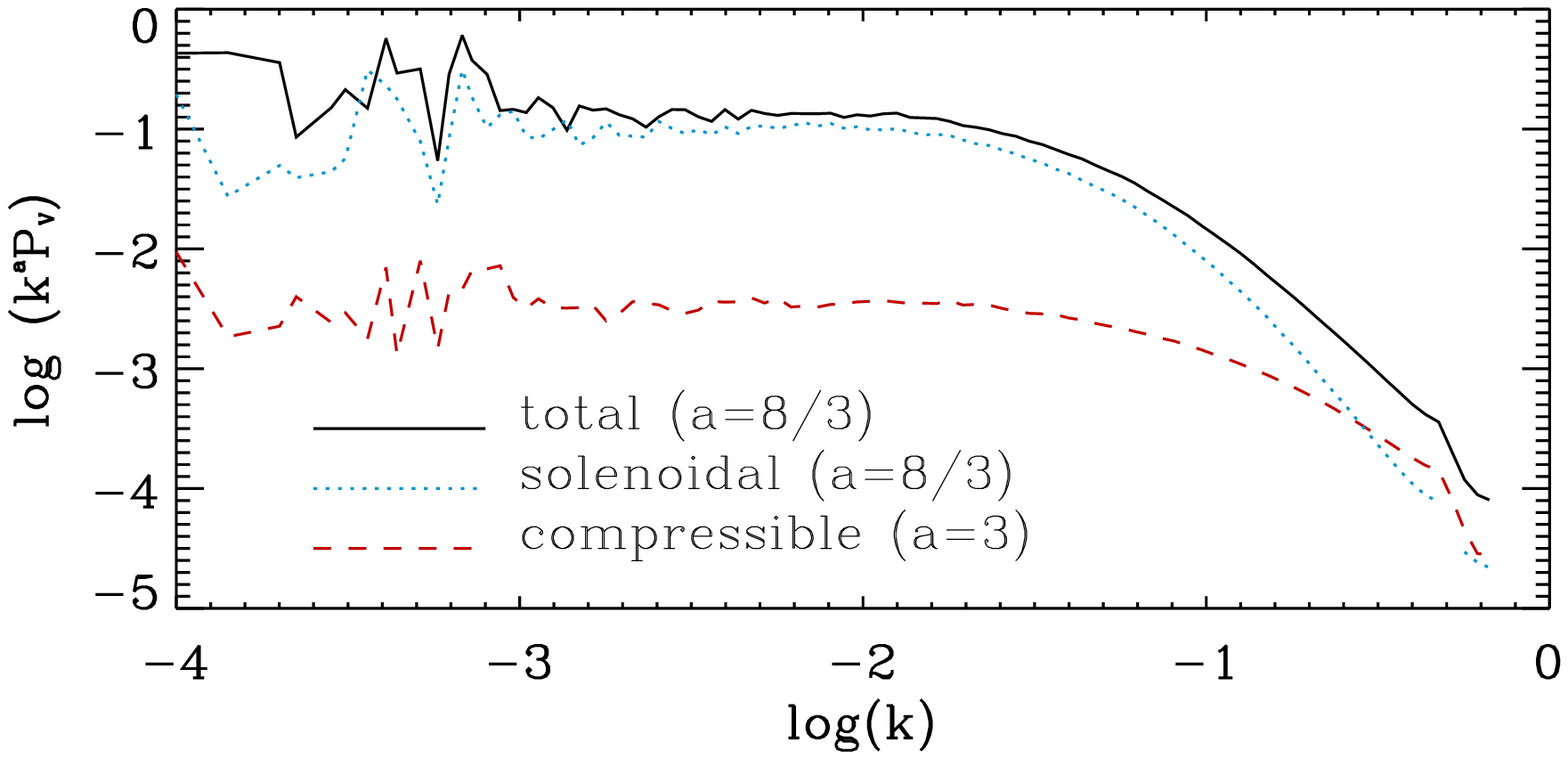}
\caption{Compensated power-spectrum of velocity (full line), solenoidal or shear component 
of velocity (dotted line), compressible component (dashed line). 
The spectra have been multiplied by $k^a$ where the value $a$ is 
indicated on the figure.}
\label{velo_sol_comp}
\end{figure}

\begin{figure}
\includegraphics[width=9cm,angle=0]{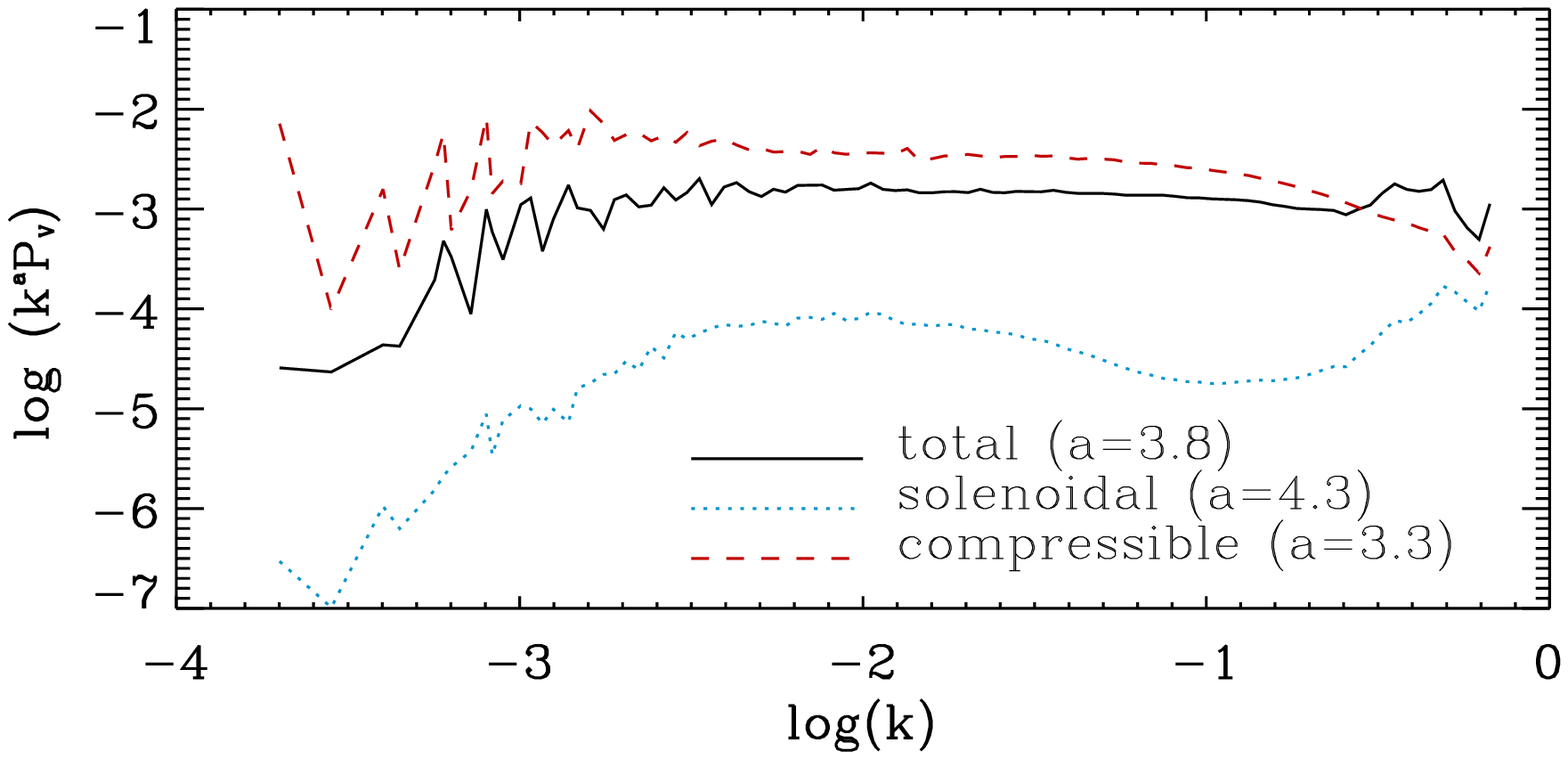}
\caption{Isothermal simulation. 
Compensated power-spectrum of velocity (full line), solenoidal or shear component 
of velocity (dotted line), compressible component (dashed line). 
The spectra have been multiplied by $k^a$ where the value $a$ is 
indicated on the figure.}
\label{velo_sol_comp_iso}
\end{figure}

\begin{figure}
\includegraphics[width=9cm,angle=0]{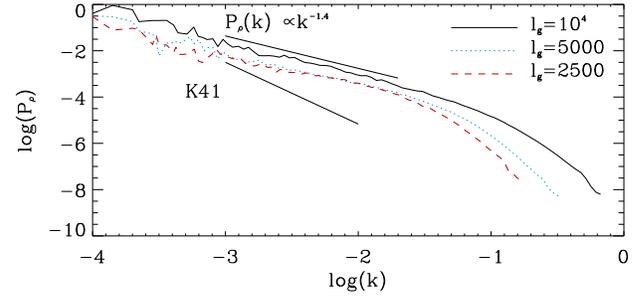}
\caption{Power-spectrum of density  for  numerical resolutions equal to 
$dx=$ $8 \times 10^{-3}$, $4 \times 10^{-3}$ and $2 \times 10^{-3}$ pc.}
\label{dens_sp2}
\end{figure}


\begin{figure}
\includegraphics[width=9cm,angle=0]{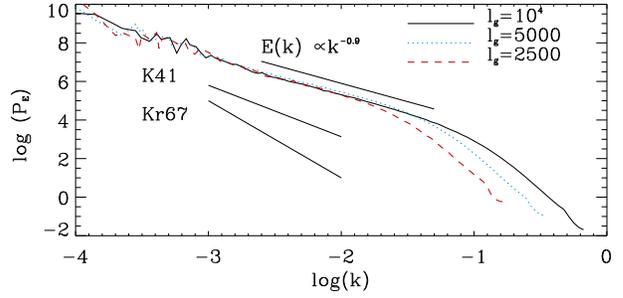}
\caption{Energy spectrum for  numerical resolutions equal to 
$dx=$ $8 \times 10^{-3}$, $4 \times 10^{-3}$ and $2 \times 10^{-3}$ pc.}
\label{ener_sp}
\end{figure}

\begin{figure}
\includegraphics[width=9cm,angle=0]{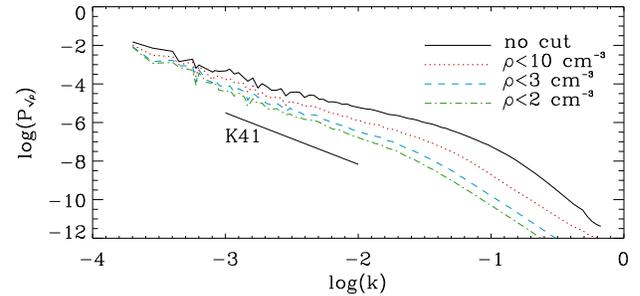}
\caption{Power-spectrum of  the square root of density field for data obtained 
by taking min$(\rho,\rho_0)$ for various values of $\rho_0$.}
\label{dens_sp_ph}
\end{figure}

\begin{figure}
\includegraphics[width=9cm,angle=0]{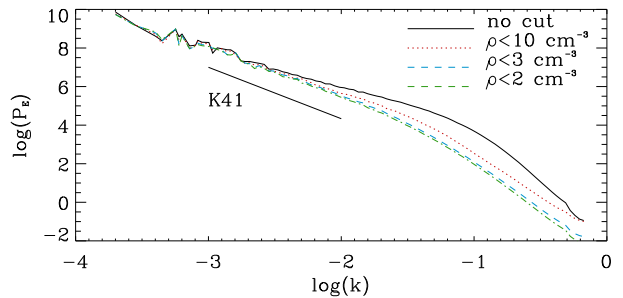}
\caption{Energy spectrum for data obtained 
by taking min$(\rho,\rho_0)$ for various values of $\rho_0$.}
\label{ener_sp_ph}
\end{figure}

\subsection{Incompressible and isothermal turbulence}

An     important diagnostic  commonly   used    in  fluid dynamics  to
characterize  the   scale-dependent   structure of the    flow  is the
power-spectrum  of the fluid variables  such  as the velocity and  the
density fields.  The power-spectrum equal  to the square of the module
of the Fourier transform, depends on all components of the wave vector
${\bf k}$.  The shell average  power-spectrum  is  the averaged of  the
power-spectrum  in spherical shells  in k-space. In the following, the
power-spectrum will refer to shell averaged power-spectrum.

In incompressible turbulence, the velocity power-spectrum, $P_V$,  is related to
the energy spectrum as $ E(k) \propto k^{D-1} P_V $, where $D$ is the
space dimension. The Kolmogorov
(Kolmogorov 1941)  prediction for the  energy spectrum of homogeneous,
3D and  incompressible turbulence is: $E(k)  \propto  k^{-5/3}$.  Note
that this corresponds  to power-spectrum proportional to $P_V \propto
k^{-11/3}$  for  3D  flows  and to  $P_V    \propto k^{-8/3}$ for  2D
flows. 
However, the behaviour of 2D incompressible turbulence is very different
from the 3D turbulence
because of enstrophy  conservation  (Kraichnan 1967). At  scales larger
than the injection scale, one expects to recover the Kolmogorov energy
spectrum whereas at scales smaller than the  injection scale, Kraichnan (1967)
predicts $k P_V \propto k ^{-3}$.

In 1D compressible turbulence,  when  shocks dominate, a  Burgers-like
spectrum $P_V \propto k^{-2}$ is expected (Elsasser \& Schamel 1976).
Passot  et al.  (1988) have  confirmed, using isothermal bidimensional
numerical simulations, that the compressible part  of the velocity field
follows the same behaviour   whereas for the solenoidal  part, the
prediction    of Kraichnan (1967) is    well verified, $k P_V \propto
k^{-3}$.  Kim \&   Ryu  (2005) recently explored  numerically  in some
details  the 3D  case.   They show   that the density   power-spectrum
($P_\rho $) for low  rms Mach number  flow is close to the Kolmogorov
power-spectrum, whereas it is much flatter at  high rms Mach number, i.e. $k^2
P_\rho \propto k^{-\alpha}$ with $\alpha  \simeq 0.5$ for an rms Mach
equal to 12.  They interpret  this result as   the consequence of  the
density  distribution    being highly   concentrated   in  sheets  and
filaments.  Flat density power-spectra have also  been observed in MHD
isothermal simulations.  Padoan et al. (2004) report $\alpha=0.25$ for
a model with  equipartition between  magnetic  and kinetic  energy and
$\alpha=0.75$ for  a  super-Alfv\'enic model  (see   also Beresnyak et
al. 2005).

\subsection{Velocity power-spectrum}
Figure~\ref{velo_ps}  shows the   velocity power-spectrum for  various
numerical resolutions,  whereas Fig.~\ref{velo_sol_comp} shows  the
compensated power-spectra  (i.e. $k^a P_V$ the value of $a$ being given in 
the figure) of the solenoidal  and  compressible components of  the
velocity field. For reference the Kolmogorov  (1941) and the Kraichnan
(1967) prediction have  been displayed in Fig.~\ref{velo_ps}.   As in Passot et
al. (1988), the compressible part  has a power-spectrum $k P_V \propto
k^{-2}$. The velocity power-spectrum,  as well  as the power-spectrum of
the solenoidal component,  appear to   be in  good agreement  with  the
Kolmogorov law, $k P_V \propto k^{-5/3}$ for $k$ between $\simeq$-3.3
and $\simeq$-1.8.

This may be at first surprising because, as recalled above, the prediction of Kraichnan for 
2D incompressible fluid is that there is an invert cascade above the energy injection 
scale, leading to the  velocity power-spectrum predicted by Kolmogorov and a
direct enstrophy cascade at scale below the injection scale leading to 
  $k P_V \propto k^{-3}$. However, this result is a consequence of the conservation 
of vorticity in incompressible or barotropic flow. When the flow is non barotropic 
the vorticity equation writes:
\begin{eqnarray}
\partial_t \omega + \nabla \times ( \omega \times v) = \nabla P \times \nabla ({ 1 \over \rho })
\label{vorticity}
\end{eqnarray}
In a radiatively cooling flow, as the interstellar atomic hydrogen, the pressure
and the density are not necessarily correlated  as already emphasized 
in section 3. This means 
that the baroclinic term on the right hand side, is not small with respect to the second term of the 
left hand side, implying
that enstrophy is not conserved. To verify this, we have computed the mean value 
of the (norm of the) baroclinic term and  of the second lhs term
 over the whole simulation. We find that the former is on average 10 times smaller
than the later. We also find that for about 7$\%$  of our grid points, the ratio 
between the two  is higher than 0.5. It seems likely to us that this constitutes a sufficiently 
large deviation from the strict enstrophy conservation to explain the deviation 
from  Kraichnan's prediction.
Unsurprisingly, we find that the higher values of the baroclinic term are obtained within 
the cold structures where the density changes much but not the pressure.

To confirm the influence of the baroclinic term, we have performed isothermal 
simulations with exactly the same setup as simulations with  cooling (same code and same 
initial and boundary conditions). In these simulations, the baroclinic term vanishes exactly. 
The compensated spectra are shown in Fig.~\ref{velo_sol_comp_iso}.
The power-spectrum of the solenoidal component is very stiff and roughly proportional to $k^{-4.4}$
whereas the compressible component has a power-spectrum proportional to about $k^{-3.2}$.
These numbers are indeed much closer to the results of Passot et al. (1988) and to the 
prediction of Kraichnan (1967). 

Finally, we note that, as revealed by Fig.~\ref{velo_sol_comp}, although  
compressible and shear modes 
are  comparable at large scales, the second dominates at small scales for ${\rm log}(k)>-3$
(solid and dotted lines are very close between $-1 > {\rm log}(k)>-3$).

\subsection{Density power-spectrum}
Figure~\ref{dens_sp2} shows the power-spectrum of the density field.
It appears to be flat since one finds $k P_\rho \propto k^{-0.4}$. 
This is reminiscent of what is observed by Kim \& Ryu (2005) 
although in the present case the strong and stiff density fluctuations 
are induced by the 2-phase nature of the flow rather than by highly supersonic 
motions.

Kim \& Ryu (2005) proposed that the flat density power-spectrum can be 
understood by the fact that the Fourier transform of a Dirac function
is independent of $k$ leading  to $P_\rho \propto k^0$ in 1D.
We believe that this explanation is also valid in the present case and 
that the density power-spectrum is  
 related to the size  distribution of the structures (Fig.~\ref{struct_reso}-\ref{struct_pix}).
Indeed since the structures  are bounded by contact discontinuities, they have stiff boundaries
 and they resemble, depending on the scales at which one is considering them, 
either to the Dirac  or Heaviside functions.
 
As a matter of fact, the end of the inertial range which occurs at about $k \simeq 0.1$ 
(about 10 cells in the simulation) corresponds to the length at which the 
 structure distribution  is influenced by the numerical diffusivity  implying 
that the number of density peaks at these scales becomes gradually smaller.  

Interestingly,  Desphande et al. (2000) using 21cm line absorption data probing 
linear scales between $ 10^{-2}$ and 3 pc, therefore comparable to the present simulation,
report density power-spectrum $k^2 P_\rho \propto k^{-\alpha}$, in two 
different regions for which  $\alpha = 0.75$ and $\alpha=0.5$.
These values are significantly flatter than the Kolmogorov index and  at least 
for the second one, are  very close to the value of 0.4 which is inferred in our simulations.

\subsection{Energy spectrum}
Figure~\ref{ener_sp} shows the power-spectrum,$P_E$,  of the quantity $\sqrt{\rho} v $.
The quantity $k P_E$ corresponds to the energy spectrum in the 
incompressible case and following Klessen et al. (2000), we call it the energy spectrum,
$E(k)$. 
It is rather different from the 
Kolmogorov prediction and significantly flatter. In the inertial range 
$E(k) \propto k^{-0.9}$, meaning that the energy is approximately equally distributed 
in the k-space. This appears to be a direct consequence of the density power-spectrum 
discussed above which relies
on the strong density fluctuations due to the 2-phase nature of the flow. 
In order to confirm  this, we have calculated the  energy spectrum 
and the power-spectrum of the square root of $\rho$ for data 
obtained by taking  min$(\rho,\rho_0)$. 
The results for various values of $\rho_0$,  namely 2, 3 and 10 cm$^{-3}$
as well as the result for the untruncated field,
 are shown in Figs.~\ref{dens_sp_ph} and~\ref{ener_sp_ph}.
When $\rho_0$ decreases,  $P_{\sqrt{\rho}}$ becomes stiffer
(note that for the untruncated field, we find  
$k P_{\sqrt{\rho}} \propto k^{-0.6}$ in the inertial range). 
The same 
behaviour is also observed for the energy spectrum which becomes closer to 
the velocity power-spectrum shown in Fig.~\ref{velo_ps}.

For $k$ larger than $\simeq 5 \times 10^{-3}$ (corresponding to a scale of about 1 pc)
 and $\rho_0 < 3$ cm$^{-3}$, the energy spectrum 
becomes significantly lower than the energy spectrum obtained for the untruncated field.
This clearly indicates that at scales smaller than about 0.5 pc,
 most of the energy is stored in the CNM  bulk motions.  In other terms, in our 
numerical experiment, 
the 2-phase nature of the flow appears to play an important r\^ole and constitutes a
significant deviation of single phase hydrodynamics at scales smaller or comparable to 
about 1 pc. 

\section{Mass spectrum} 
As in paper I, we extract the structures by a 
simple clipping algorithm. Since the cold structures are connected 
to the surrounding warm gas by stiff thermal fronts such simple 
criteria has a clear physical meaning. The threshold is equal to 
$30$ cm$^{-3}$ which lies in the  thermally unstable regime
(note that we have verified that taking $10$ cm$^{-3}$ instead does not 
change the results significantly).

\subsection{Numerical result}

\begin{figure}
\includegraphics[width=9cm,angle=0]{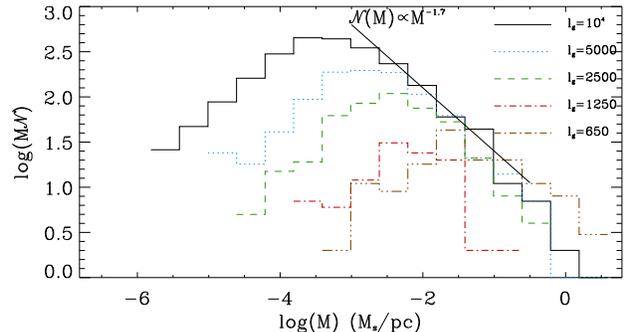}
\caption{Number of structures per logarithmic interval of mass 
(in solar mass per parsec). Various numerical resolutions are shown. }
\label{struct_reso}
\end{figure}

\begin{figure}
\includegraphics[width=9cm,angle=0]{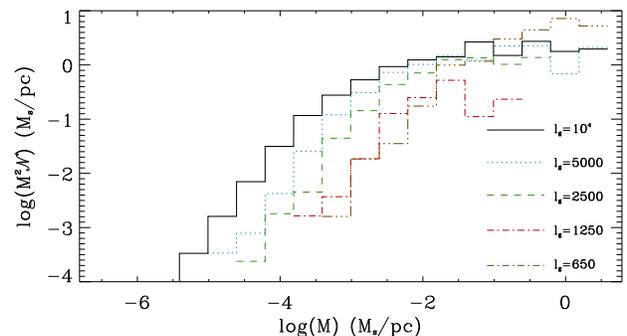}
\caption{Total mass within CNM structures per logarithmic interval of mass.}
\label{struct_reso_pond}
\end{figure}

\begin{figure}
\includegraphics[width=9cm,angle=0]{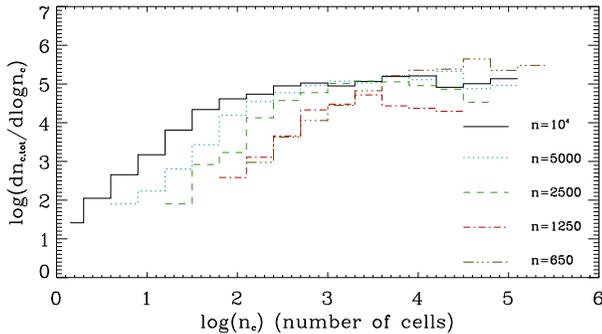}
\caption{Total number of cells per logarithmic interval of 
cells number 
 (expressed in cells corresponding to the highest resolution run). }
\label{struct_pix}
\end{figure}

Figure~\ref{struct_reso}  shows   the    number   of  structures, $n_s$,   per
logarithmic mass interval, $d  n _s/ d{\rm log} M$
(note that since the simulations are 2D, the mass is expressed  in solar mass  per
parsec).  The largest structures  have a mass of  about one solar mass
per  parsec whereas  the mass   of  the smallest  structures  is about
$\simeq  10^{-5}$ solar mass per  parsec. The number of structures, 
${\cal N}(M)$, 
 decreases with mass, and  follows
approximately: $d n_s /  d{\rm log} M \propto 1 /M^{\beta-1}$ with $\beta \simeq 1.7$, 
until ${\rm log}(M) \simeq -3$. This implies that the number of structures
per mass interval  is $d n_s / d M = {\cal N}(M)  \propto 1 / M^\beta$.
 
For ${\rm log}(M) \le -3$, 
 $d n_s/ d {\rm log} M $   increases with mass.  
This
part of   the distribution is  most  likely affected by  the numerical
resolution.   This is confirmed  by the  fact that  the numbers  of low mass
objects decreases  with the numerical resolution. Interestingly enough,
the number of massive structures becomes independent of the resolution
for $dx \le 10^{-2}$ pc (see solid, dotted and  dashed curves). 
This indicates that numerical  convergence is reached for the
large structures  (${\rm log}(M)  \ga -3$). It  is  however  clear that  no
numerical convergence    has yet been reached  for the  smallest
structures.
We also note that the number of structures more massive than 
$\simeq 0.3$  M$_s$ pc$^{-1}$, appears to be smaller than the value 
expected from the power-law fitted above. This is most likely due to the 
finite size of the numerical experiment.

For ${\rm log}(M) \ge -3$, 
the  total mass contained in a logarithmic interval of mass, follows
approximately 
$dM_{\rm tot} / d {\rm log} \propto M ^2 {\cal N} \simeq M ^{2-\beta} \simeq M^{0.25-0.3}$ which indicates 
that, although the mass of CNM is, in principle, dominated by the large 
clouds, the index of the power-law is nevertheless very shallow. 
This is confirmed by 
Fig.~\ref{struct_reso_pond}  which shows   $dM_{\rm tot} / d {\rm log}$
 as a function of ${\rm log}(M)$. It is seen that  the mass in large structures is
only slightly larger than  the mass in  hundred  to thousand times  smaller
structures.  This indicates   that    the description of   the   small
structures is  very  important in order to achieve a fair description of the  flow and
its   dynamics. 

Complementary  information is   given in Fig.~\ref{struct_pix}   which
shows  the  total number  of  cells, $d n_{c, {\rm tot}} / d {\rm log} n_c$,
 per logarithmic interval of computational cells number, $n_c$, 
contained in the structure  (expressed
in cells  of the  highest resolution  run). The  physical size  of the
structures is obtained  by taking the square root  of the cell  number
and  multiplying by $2 \times 10^{-3}$ pc which  is the numerical resolution of the
10000$^2$ cells run.  The size  at which the structure
number starts to decrease because of insufficient numerical resolution
is about 10 ($\simeq  \sqrt{n_c}$ for $n_c\simeq 100$). The length of the
bigger  structures is  roughly  300  cells  ($\simeq  \sqrt{n_c}$   for
$n_c\simeq 10^5$) which corresponds to about 0.6 pc.

\subsection{Influence of  thermal conduction}

\begin{figure}
\includegraphics[width=9cm,angle=0]{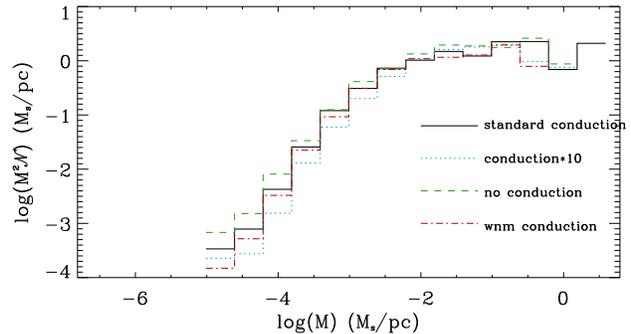}
\caption{Same as Fig.~\ref{struct_reso_pond} except that the influence
of various conductivity is shown.}
\label{struct_reso_pond_cond}
\end{figure}

The importance of thermal conductivity has been recently pointed out 
by Koyama \& Inutsuka (2004). 
They show explicit cases for which the number of structures which forms 
do depend on the conductivity.
In order to investigate the  exact influence of the thermal conductivity on the 
structures, we have performed various runs in which its value has been changed.
The results are presented in Fig.~\ref{struct_reso_pond_cond}.
Four cases have been explored. The full black line is for standard ISM conduction, the 
dashed green line is when no explicit conduction  is taken into account and the 
dotted blue line is for a conduction 10 times the standard ISM conditions. 
Finally, in order to determine the importance of the dependence in temperature of the 
conduction, we also perform a run in which the conduction does not vary with 
temperature and is equal to the conduction at $T=8000$ K. The corresponding run 
is shown by the dot-dashed red line.
The number of structures more massive than $\simeq 10^{-3} $ solar mass per parsec
appears to be independent of thermal conduction. There is a clear influence 
of thermal conduction on smaller structures, larger conduction runs produce 
less small structures than smaller conduction runs. This is qualitatively
in good agreement with the trend expected from  Field's linear analysis that 
shows that the conductivity prevents the formation of structures smaller than the 
Field length.
However, the discrepancy between the different curves is not very large. 
We therefore conclude that  thermal conduction, although not 
insignificant, does not appear to play a dominant r\^ole.

\section{Theoretical derivation of the mass spectrum}
Cloud mass spectrum is an important feature of the flow. It would be useful to 
have a theoretical model to explain the numerical results. Here, we propose
an approach inspired from the Press \& Schecter (1974) formalism (see also Padmanabhan 1993).

\subsection{Principle and justification of the analysis}
The starting point of the Press \& schecter (1974) formalism is to consider the 
density spectrum of the initial fluctuation that eventually lead to  
galaxies. 
They  consider 
the density field smoothed at a given scale, $R$, by a window function $W$ and 
assume that the probability of finding a contrast density $\delta  = \rho / \overline{\rho} - 1$,
 $\overline{\rho}$ being the mean density, is
\begin{eqnarray}
P_R (\delta ) = {1 \over \sqrt{2 \pi} \sigma(R) } \exp\left( {- \delta ^2 \over 2 \sigma(R) ^2 } \right)
\label{prob_rho}
\end{eqnarray}
where 
\begin{eqnarray}
\sigma ^2 (R) = \int \widetilde{\delta}^2 (k) W _k ^2(R) d^3k,
\label{sigma}
\end{eqnarray}
 (see e.g. Padmanabhan 1993, chapter 5).
In this last expression $\widetilde{\delta}^2 (k)$ is the power spectrum of $\delta$.

\begin{figure}
\includegraphics[width=8cm]{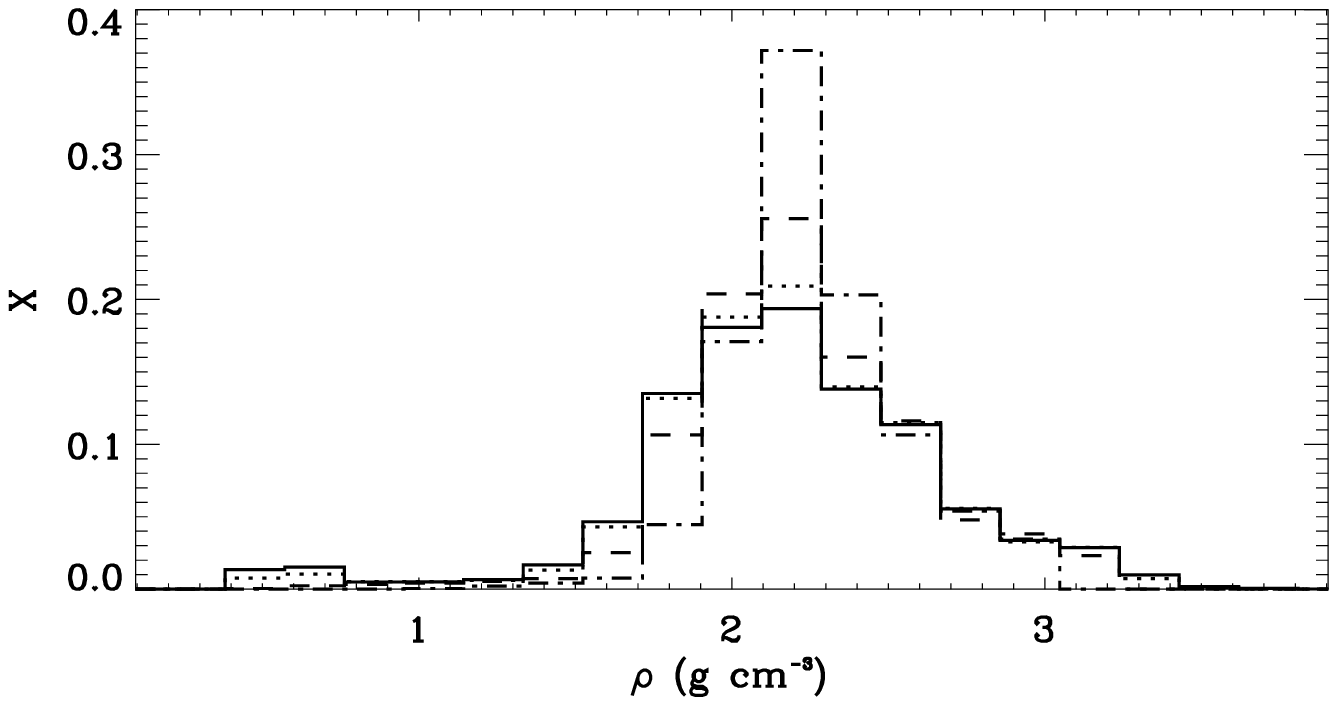}
\caption{Isothermal simulation. Density field smoothed at scale $R=b \times dx$, for 
various values of $b$. Solid line is $b=100$, dashed line is $b=200$, dashed line is 
$b=500$, whereas dot-dashed line is $b=1000$.}
\label{fluctuation_iso}
\end{figure}

\begin{figure}
\includegraphics[width=8cm]{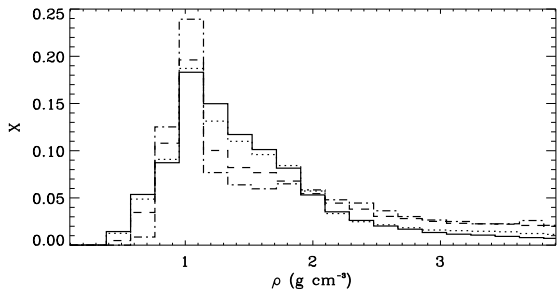}
\caption{2-phase simulation. Density field smoothed at scale $R=b \times dx$, for 
various values of $b$. Solid line is $b=100$, dashed line is $b=200$, dashed line is 
$b=500$, whereas dot-dashed line is $b=1000$. }
\label{fluctuation_cool}
\end{figure}

Therefore, the first question that must be addressed is whether this assumption is justified in 
the case of the atomic hydrogen. More precisely, since the CNM structures, are the final product of 
density fluctuations arising in the WNM, the question which has to be addressed is whether the  
spectrum  of the density fluctuations smoothed at a scale $R$, is reasonably described by 
the probability stated by Eq.~\ref{prob_rho} and what is the value of $\sigma (R)$. 
One difficulty to address this question is that in 
the multiphase simulations presented previously, the WNM and the CNM are mixed together. 
They are therefore not representative of the initial state of the WNM  which leads to the CNM 
structures. On the other hand,  the early times of the simulation reflect the numerical setup rather than the 
initial conditions. 
To overcome this difficulty, we consider the simulation with cooling presented above
but in which we have removed the CNM by taking min$(\rho,\rho_0)$, with $\rho_0=3$ cm$^{-3}$
(see section~5.4). As  shown by Fig.~\ref{dens_sp_ph}, the power-spectrum of $\sqrt{\rho}$
is nearly Kolmogorov. We have checked that the power-spectrum of the density field is indeed
nearly Kolmogorov as well.  Another point is that the power-spectrum of the density in 
the isothermal simulation performed in Sect.~5.2 
(let us remember that this simulation has exactly the same initial, boundary and forcing conditions than 
the other runs performed with cooling) is indeed very close to the Kolmogorov spectrum. The reason
is that the isothermal simulation is subsonic (as for the simulation with cooling, the rms velocity is 
about half the sound speed of the WNM). Since in the ISM, the WNM is likely to be more or 
less isothermal until the thermal pressure is triggered to a value for which WNM is thermally 
unstable, this should constitute a reasonable description of WNM before the thermal transition occurs.
Therefore,  we assume that  $\widetilde{\delta} ^2  (k) \propto k ^{-n}$, with $n \simeq 8/3$.

To calculate the width of the distribution of the  density fluctuations for the case 
of the turbulent  WNM, we choose
the simple window function sharply truncated in k-space,  $W _k(R) = \theta(R^{-1} -k) $ where 
$\theta(z) =1 $ if $z \ge 0$ and 0 otherwise.
With Eq.~\ref{sigma}, it is easy to see that the integral diverges for small $k$. This is a 
natural consequence of the well known property of the Kolmogorov spectrum that most of the 
energy is on the large scales.We therefore have to introduce a lower value in k-space, which corresponds
to an upper limit of the spatial scales, $L_c$.  In the case of the simulation, this is naturally given by 
the size of the computational box whereas in the ISM,  this would be the energy injection scale. 
We have:
\begin{eqnarray}
\sigma^2(R)  = \int _{1/L_c} ^{1/R} C k k^{-n} dk = { C \over n -2 } \Big( \big( {1 \over L_c } \big)^{2-n} 
- \big({1 \over R } \big) ^{2-n}  \Big),
\label{sigma_calc}
\end{eqnarray}
where $C$ is a constant.

The question is then, are the density fluctuations reasonably described by Eqs.~\ref{prob_rho}
and~\ref{sigma_calc} ?
To answer this question, we have smoothed the density field of the isothermal 
and cooling simulations using the  simple window function, 
$W_b(\rho_{i,j}) = 1 / b ^2 \sum _{  \vert i-m \vert \le b , \vert j-n \vert \le b} \rho_{m,n} $, 
for various values of $b$.
Figures~\ref{fluctuation_iso} and~\ref{fluctuation_cool} show the results in the isothermal case  and in the  case with cooling
(since the isothermal simulation has 5000$^2$ cells, the values of $b$ are 2 times lower 
than in the case with cooling).
It is seen that, in the isothermal case, the probability distribution is reasonably Gaussian. As expected
from Eq.~\ref{sigma_calc}, $\sigma$ appears to be a decreasing function of $b=R/dx$. 
In the case with cooling, the situation is slightly less clear because 
the shape is a little skewed towards the 
large densities.  This is  a consequence of the fact that a fraction of the gas is 
thermally unstable and in transition towards CNM or is already CNM.  
This dense gas is therefore on the process of 
condensation or already condensed and should not be considered in the pdf of density fluctuations of WNM. 

Beyond this qualitative agreement, we present in Appendix A a more quantitative estimate 
that confirms the validity of Eq.~\ref{sigma_calc}. 
 Thus it appears that the assumption 
of the distribution of the density fluctuations being given by Eqs.~\ref{prob_rho} and
~\ref{sigma_calc} is sufficiently  justified.

\subsection{Mass function}
To calculate the mass function, we can now follow the same procedure as for the cosmological 
calculation.  We note, however, two important differences. 
First, since we assume statistical stationarity, there is no time dependence in our analysis. 
Second, the collapse is not due to gravity but to thermal instability. 

Here, since our primary goal is to understand  the simulation results, we consider the 2D case.  
Thus, for a perturbation of size, $L$, the  final mass of the structures is about $M(L) = \overline{\rho} L ^2$, 
Eq.~\ref{sigma_calc} is equivalent to
\begin{eqnarray}
\sigma^2(M)  =  { C \over n -2 } \Big(  M (L_c) ^{(n-2)/2} 
- M^{(n-2)/2}  \Big) \overline{\rho}^{(2-n)/2},
\label{sigma_calc2}
\end{eqnarray}
 where $M=M(R)$.

The fraction of objects having a mass larger than $M$ is thus given by:
\begin{eqnarray}
F(M) = \int _0 ^\infty P_R(\delta) s(\delta) d \delta,  
\label{fraction}
\end{eqnarray}
where $s$ is the selection function
\footnote{We note that in cosmology this expression is known to be inconsistent because 
in the cosmological case, the integral of $dF= f(m) dm$ from 0 to $\infty$ should be equal to 1. 
This problem has been solved by the Excursion set theory (Bond et al. 1991). The results for the 
shape of the mass spectrum is however independent of this and we will not consider this 
problem further at this stage. }. 
The selection function, $s$, determines the fluctuations
that will eventually lead to the formation of a CNM structure.  
For simplicity, we will assume that once the density 
reaches a threshold $\delta_c$, it leads to a thermal collapse and to the formation of 
a CNM structure. Although this assumption is reasonable given the physical characteristics of thermal instability, 
we note that various improvements
could certainly be possible, as for example taking into account the stabilizing r\^ole of the 
turbulence within the perturbation (in that case $s$ would also depend  on the velocity 
dispersion).
Thus we have
\begin{eqnarray}
F(M) = \int _{\delta_c} ^\infty P_R(\delta)  d \delta.   
\label{fraction2}
\end{eqnarray}

The  mass function, ${\cal N}(M)$,  is then simply given by $-\partial F / \partial M / (M/\overline{\rho}) $.
We have \footnote{The Excursion set theory would give the same result multiplied by a factor 2.}
\begin{eqnarray}
{\cal N}(M) dM = - {1 \over \sqrt{2 \pi}} { \overline{\rho} \over M } {\delta_c \over \sigma ^2} {d \sigma \over d M} 
\exp \left( { - \delta_c^2 \over 2 \sigma^2} \right) dM.  
\label{mass_func}
\end{eqnarray}

With Eq.~\ref{sigma_calc2}, we get
\begin{eqnarray}
{d \sigma \over d M} = - \sqrt{C} \overline{\rho}^{1/2-n/4} &\times& \\ 
 {M ^{(n-2)/4} \over 2 M }  &\times&  \sqrt{ {(n-2)  \over  \Big(  M (L_c) ^{(n-2)/2} 
- M^{(n-2)/2}  \Big) } }. 
\label{dsig_dM}
\nonumber
\end{eqnarray}
This leads to:
\begin{eqnarray}
{\cal N}(M) dM \propto  M ^{(n-2)/2 - 2} {1 \over \sigma ^3}  \exp( - \delta_c^2 / (2 \sigma^2) ) dM  
\label{mass_func2}
\end{eqnarray}
Therefore for small masses, i.e. mass smaller than $M(L_c)$, we obtain that ${\cal N}(M) \propto M^{(n-2)/2 -2}$
with $n \simeq 8/3$, we get ${\cal N}(M) \propto M ^{-5/3}$. We note that $5/3 \simeq 1.666$ is rather close to the 
value $1.7$ obtained in the simulation. For a mass close to $M(L_c)$, there is an exponential cutoff which is 
due to the cut, $L_c$ introduced at large scales. Altogether these results are similar to the results obtained 
in the cosmological case. An important difference however lies in the exponent of the 
mass spectrum which is less stiff. 
This is a consequence of the fact that in a turbulent  flow, most of the energy is at
 large scales implying   more large scale fluctuations.

\subsection{The 3D case: similarity with the CO clumps}
This analysis can be straightforwardly applied to the 3D case. The only differences
are that $n=11/3$ (since the simulations presented here are bidimensional, it is 
worth to remember that  Kim \& Ryu (2005)  do find that in transonic isothermal simulations, 
the density power spectrum is nearly Kolmogorov) and $M = \overline{\rho} R^3$. We obtain:
\begin{eqnarray}
{d \sigma \over d M} \propto -  {M ^{(n-3)/6} \over 3 M } \sqrt{ {(n-3)  \over  \Big(  M (L_c) ^{(n-3)/3} 
- M^{(n-3)/3}  \Big) } },
\label{dsig_dM_3D}
\end{eqnarray}
and
\begin{eqnarray}
{\cal N}(M) dM \propto  M ^{(n-3)/3 - 2} {1 \over \sigma ^3}  \exp( - \delta_c^2 / (2 \sigma^2) ) dM.  
\label{mass_func2_3D}
\end{eqnarray}
The index of the power law of the mass spectrum in the 3D case and 
for masses small compared to  $M(L_c)$,  is about $(n-3)/3 \simeq 1.78$, i.e. slightly stiffer 
than the value inferred in the 2D case. 

Interestingly, we note that the value inferred here, turns out to be similar to the index of the mass
spectrum observed for the CO clumps of masses  between 10$^4$ solar mass  and one Jupiter mass 
(Kramer et al. 1998, Heithausen et al. 1998). 

Although it cannot be excluded that this is a pure coincidence, one possibility is that the 
origin of the CO clumps is indeed rooted in the atomic gas.  The mass distribution of the 
CO clumps may therefore be determined during the atomic phase before the gas becomes molecular.

\section{Conclusion}
We have presented high resolution numerical simulations aiming to describe 
a turbulent interstellar atomic flow. The high resolution ($2 \times 10^{-3}$ pc) 
provides a good description (although not sufficient to reach numerical 
convergence)  of the flow close to  the spatial scales of the 
smallest structures observed  in HI. We confirm the main results obtained in 
paper I. The flow is very fragmented and the phases are tightly interwoven.
The CNM structures are connected to the WNM by stiff thermal fronts and are 
locally in near pressure equilibrium with the surrounding WNM. Altogether, this is 
reminiscent of the classical 2-phase model in spite of the fact that the flow is dynamical
and turbulent.

We find that small scale structures, either diffuse or very dense,  are naturally produced 
in a turbulent 2-phase medium. Whereas the former are simply the tail of the structure mass spectrum
induced by turbulence, the latter, which are somehow reminiscent of the TSAS observed in the atomic gas, 
are transient events produced by supersonic collisions between CNM fragments.  

We characterize the flow by studying various statistical quantities, namely 
the mass distribution of the CNM structures, the velocity and density power-spectrum and the 
energy spectrum, paying special attention to the numerical resolution. 
We stress that the present problem requires a lot of numerical resolutions to 
obtain reliable results. Typically at least 5000$^2$ to 10000$^2$ cells are needed and even so, 
no strict convergence is achieved. 

Our main results are as follows.

For  structures of size larger than about 10 
grid cells, for which numerical diffusivity is not too important, 
we find a mass spectrum, ${\cal N}(M) \propto M^{-1.7}$. In order to 
explain this result, we have carried out a calculation based on 
Press \& Schecter (1974) formalism. 
Our theory predicts 
${\cal N}(M) \propto M^{-5/3}$ in 2D and ${\cal N}(M) \propto M^{-16/9}$ in 3D.
One of the main assumptions is that the CNM clumps are due to density 
fluctuations within WNM, which behaves as a sub to transonic nearly isothermal 
gas. We note that these mass spectra are similar to the mass spectrum 
inferred observationally for CO clumps. This is compatible with the origin of
 CO clumps being rooted  in the very diffuse atomic gas although this does 
not, by any means, constitute a proof.

Like in highly supersonic isothermal  flows, the density power-spectrum is rather flat. 
However, 
the structures being pressure-bounded and not bounded by shocks as in supersonic
isothermal simulations, they are  long-living objects. 
Unlike supersonic isothermal simulations, the velocity power-spectrum follows the Kolmogorov prediction
and is dominated by its solenoidal part. 
 At scales larger than about 1 pc, most of the energy is in the WNM whereas at scales 
smaller than about 1 pc, it is mostly in the CNM. Indeed, 
the energy is  nearly equally distributed in k-space for scales ranging 
between about 2 and $2 \times 10^{-2}$ pc where numerical diffusion becomes dominant.
This behaviour is due to the density structure of the flow that presents 
relatively flat density power-spectrum, indicating that  in a thermally bistable flow like 
the one we studied, the kinetic energy 
at scale below 1 pc, appears to be dominated by the 
translational motions of the CNM fragments.

We also study the influence of the 
thermal conduction by performing various runs with different thermal conductivities and 
find a modest influence, mainly on the small structures. 

In a companion paper (paper III) we  further characterize the simulation results by 
studying line of sight statistics and the physical properties of the CNM clouds. 
We also calculate synthetic HI spectra and reach the  conclusion that, 
although the lines of sight from which they are calculated are very complex, showing 
in some cases several structures, the spectra are relatively smooth and simply broadened 
by the complex motions along the line of sight.

Altogether these results suggest  that the turbulence which takes place in the 
neutral atomic interstellar gas 
is very  different from the turbulence which takes place in an isothermal or polytropic gas. 
 One important question that remains to be 
investigated is how these conclusions will change in a more realistic framework of 
3D flows. To answer this question 3D simulations have been performed and will be 
presented elsewhere (Audit \& Hennebelle 2007).

\begin{acknowledgements}
We acknowledge  the support of  the CEA computing center,  CCRT, where
all the  simulations where  carried out. PH is most grateful to John
Scalo for stimulating exchanges on the interstellar turbulence as well as 
insightful comments. We thank Snezana Stanimirovi\'c, the referee, for 
constructive and interesting comments which have improved 
the paper significantly.
\end{acknowledgements}

\appendix

\section{Further measurement of the pdf of density fluctuations}

\begin{figure}
\includegraphics[width=8cm]{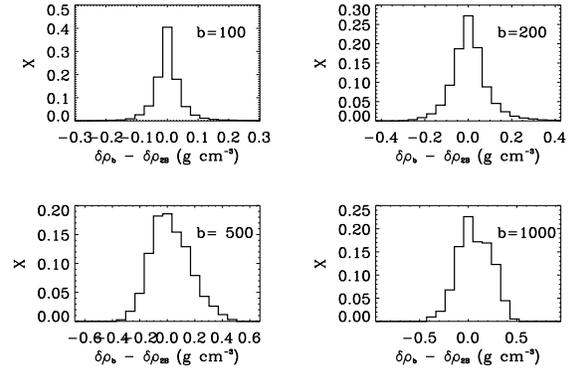}
\caption{Isothermal simulation. Density field smoothed by the 
window function $W_b - W_{2b}$ for various value of $b$. The range of the 
x-axis being proportional to $b^{n-2}$, it appears that the length of these 
distribution is reasonably gaussian and roughly proportional to $b^{n-2}$.}
\label{fluctuation2_iso}
\end{figure}

\begin{figure}
\includegraphics[width=8cm]{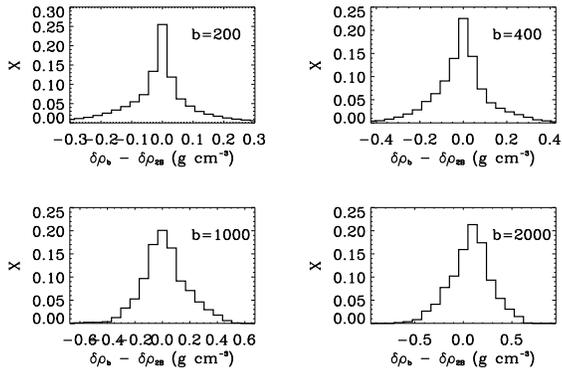}
\caption{Same as Fig.~\ref{fluctuation2_iso} for the 2-phase
simulation in which  
the density field has been truncated by taking min$(\rho,\rho_0)$.}
\label{fluctuation2_cool}
\end{figure}

In order to measure more accurately the dependence of $\sigma$ (Eq.~\ref{sigma_calc}) on $R$,
  we  define  $\delta \rho _b - \delta \rho_ {2b} = W _b (\rho) - W _{2b} (\rho) $, i.e. the 
density field smoothed by the window function  $W_b - W_{2b}$. 
With this window function, one finds that $\sigma(R) = C/(n-2) \times (2^{n-2} -1) R ^{n-2}$.
In particular, it is independent of  $L_c$ and is proportional to $R^{n-2} = R^{2/3}$
for $n=8/3$.

Figures~\ref{fluctuation2_iso} and~\ref{fluctuation2_cool} display the distribution 
of $\delta \rho _b - \delta \rho_ {2b}$
for various values of $b$ for the isothermal and  the 2-phase simulations. 
The range of the x-axis is proportional to $b^{n-2}$. As can be seen, the
distributions are reasonably Gaussian and the  width 
is broadly proportional to $b^{n-2}$ for both cases except for small $b$ 
in the isothermal simulation.


\begin{thebibliography}{99}



\bibitem{audit-hennebelle}
Audit, E., Hennebelle, P., 
{2005, A\&A 433, 1 (paper I)}

\bibitem{aud-hen2}
Audit, E., Hennebelle, P., 
{2007, {\it in preparation} }

\bibitem{beresnyak}
Beresnyak, A., Lazarian, A., Cho, J., 
{\ 2005, ApJ 624L, 93}

\bibitem{bond}
Bond, J., Cole, S., Efstathiou, G., Kaiser, N., 
{\ 1991, ApJ 379, 440}


\bibitem{braun}
Braun, R., Kanekar, N., 
{\ 2005, A\&A 436L, 53}


\bibitem{crovisier}
Crovisier, J., 
{\ 1981, A\&A 94, 162}

\bibitem{de avillez}
de Avillez, M., Breitschwerdt, D., 
{\ 2005, A\&A 436, 585}


\bibitem{dib}
Dib, S., Burkert, A., 
{\ 2005, A\&A 630, 238}

\bibitem{dickey}
Dickey, J., Lockman, F., 
{\ 1990, ARA\&A, 28, 215}



\bibitem{field}
Field, G.,
{\ 1965, ApJ  142, 531}

\bibitem{fieldgold}
Field, G., Goldsmith, D., Habing, H.,
{\ 1969, ApJ Lett 155, 149}


\bibitem{folini}
Folini, D., Walder, R., 
{\ 2006, A\&A, {\it in press}, astro-ph 0606753}

\bibitem{gazol01}
Gazol, A., V\'azquez-Semadeni, E., S\'anchez-Salcedo, F., Scalo, J.,
{\ 2001, ApJ 557, L124}


\bibitem{gazol05}
Gazol, A., V\'azquez-Semadeni, E., Kim, J.,
{\ 2005, ApJ 630, 911}




\bibitem{heiles}
Heiles, C.,
{\ 1997, ApJ 481, 193}


\bibitem{heilestroland}
Heiles, C., Troland, T.,
{\ 2003, ApJ 586, 1067}


\bibitem{heilestroland2}
Heiles, C., Troland, T.,
{\ 2005, ApJ 624, 773}

\bibitem{heithausen}
Heithausen, A., Bensch, F., Stutzki, J., Falgarone, F., Panis, J.-F., 
{\ 1998, A\&A 331, L65}

\bibitem{heitsch05}
Heitsch, F., Burkert, A., Hartmann, L., Slyz, A., Devriendt, J., 
{\ 2005, ApJ 633, 113}

\bibitem{heitsch06}
Heitsch, F., Slyz, A., Devriendt, J.,  Hartmann, L., Burkert, A.,  
{\ 2006, ApJ 648, 1052 }


\bibitem{hen1}
Hennebelle, P., P\'erault, M., 
{\ 1999, A\&A 351, 309}

\bibitem{hen2}
Hennebelle, P., P\'erault, M., 
{\ 2000, A\&A 359, 1124}


\bibitem{hen-pas}
Hennebelle, P., Passot, T., 
{\ 2006, A\&A 448, 1083}





\bibitem{hen-aud-miv}
Hennebelle, P., Audit, E., Miville-Desch\`enes, M.-A.,
{\ 2006, A\&A, {\it submitted} (paper III)}



\bibitem{kim_ryu}
Kim, J., Ryu, D., 
{\ 2005, ApJ 630, L45}


\bibitem{klessen}
Klessen, R., Heitsch, F., Mac Low, M.-M.,
{\ 2000, ApJ 535, 887}

\bibitem{kramer}
Kramer, C., Stutzki, J., Rohrig, R., Corneliussen, U., 
{\ 1998, A\&A 329, 249}

\bibitem{kolmogorov}
Kolmogorov, A.N.
{\ 1941, Proc. R. Soc. London Ser. A 434,  Reprinted in 1991}


\bibitem{koyoma1}
Koyama, H., Inutsuka, S., 
{\ 2000, ApJ 532, 980}


\bibitem{koyoma2}
Koyama, H., Inutsuka, S., 
{\ 2002, ApJ 564, L97}


\bibitem{koyoma3}
Koyama, H., Inutsuka, S., 
{\ 2004, ApJ 602L, 25}

\bibitem{koyoma6}
Koyama, H., Inutsuka, S., 
{\ 2006, ApJ  {\it in press}, astro-ph/0605528}



\bibitem{kraichnan}
Kraichnan, R.H., 
{\ 1967, Phys. Fluid 8, 1385}

\bibitem{kritsuk1}
Kritsuk, A. G., Norman, M. L.,  
{\ 2002, ApJ 569, L127}



\bibitem{kulk}
Kulkarni, S.R., Heiles, C.,
{\ 1987, {\it  Interstellar processes}, ed. Hollenbach D., 
Thronson H. (Reidel)}


\bibitem{miville}
Miville-Desch\^enes, M.-A., Joncas, G., Falgarone, E., Boulanger, F.,
{\ 2003, A\&A 411, 109 } 


\bibitem{padmanabhan}
Padmanabhan, T., 
{\ 1993, {\it Structure formation in the universe}, Cambridge university press}


\bibitem{padoan}
Padoan, P., Jimenez, R., Juvela, M., Nordlund, A., 
{\ 2004, ApJ 604, 49}


\bibitem{passot}
Passot, T., Pouquet, A., Woodward, P., 
{\ 1988, A\&A, 197, 228}


\bibitem{passot-vazquez}
Passot, T., Vazquez-Semadeni, E., 
{\ 1998, Phys. Rev. E., 58, 4501}


\bibitem{Pens}
Penston, M., Brown, F.,
{\ 1970, MNRAS 150, 373}

\bibitem{piontek}
Piontek, R., Ostriker, E., 
{\ 2004, ApJ 601, 905 } 


\bibitem{press}
Press, W., Schecter, P., 
{\ 1974, ApJ 187, 425}


\bibitem{Sanchez}
S\'anchez-Salcedo, F. J., V\'azquez-Semadeni, E., Gazol, A.,
{\ 2002, ApJ 577, 768}


\bibitem{scalo}
Scalo, J., Vazquez-Semadeni, E., Chappell, D., Passot, T., 
{\ 1998, 504, 835}


\bibitem{stanimirovic}
Stanimirovi\'c, S., Heiles, C., 
{\ 2005, ApJ  631, 371 }

\bibitem{toro}
Toro, E., 
{\ 1997,  {\it Riemann solvers and numerical methods for fluid dynamics}, 
(Springer) }

\bibitem{vazquez-semadeni}
V\'azquez-Semadeni, E., Ryu, D., Passot, T., Gonz\'alez, R., Gazol, A., 
{\ 2006, ApJ 643, 245}

\bibitem{vishniac}
Vishniac, E., 
{\ 1994, ApJ, 428, 186}

\bibitem{walder96}
Walder, R., Folini, D., 
{\ 1996, A\&A, 315, 265}

\bibitem{walder98}
Walder, R., Folini, D., 
{\ 1998, A\&A, 330L, 23}


\bibitem{wolfire95}
Wolfire, M.G.,  Hollenbach, D., McKee, C.F.,
{\ 1995, ApJ 443, 152}

\bibitem{wolfire03}
Wolfire, M.G.,  Hollenbach, D., McKee, C.F.,
{\ 2003, ApJ 587, 278}

\bibitem{Zel}
Zel'dovich, Y., Pikel'ner, S.,
{\ 1969, JETP 29, 170}


\end{thebibliography}
\end{document}